\documentclass[aps,twocolumn,prl,preprintnumbers,amsmath,amssymb,superscriptaddress,floatfix]{revtex4}
\usepackage[pdftex]{graphicx}
\usepackage{amsmath}

\usepackage{hyperref}
\begin{document}

\title{Signatures of Ultrafast Reversal of Excitonic Order in Ta\textsubscript{2}NiSe\textsubscript{5}}

\author{H. Ning}
\affiliation{Institute for Quantum Information and Matter, California Institute of Technology, Pasadena, CA 91125}
\affiliation{Department of Physics, California Institute of Technology, Pasadena, CA 91125}

\author{O. Mehio}
\affiliation{Institute for Quantum Information and Matter, California Institute of Technology, Pasadena, CA 91125}
\affiliation{Department of Physics, California Institute of Technology, Pasadena, CA 91125}

\author{M. Buchhold}
\affiliation{Institute for Quantum Information and Matter, California Institute of Technology, Pasadena, CA 91125}
\affiliation{Department of Physics, California Institute of Technology, Pasadena, CA 91125}

\author{T. Kurumaji}
\affiliation{Department of Physics, Massachusetts Institute of Technology, 77 Massachusetts Ave., Cambridge, MA 02139}

\author{G. Refael}
\affiliation{Institute for Quantum Information and Matter, California Institute of Technology, Pasadena, CA 91125}
\affiliation{Department of Physics, California Institute of Technology, Pasadena, CA 91125}

\author{J. G. Checkelsky}
\affiliation{Department of Physics, Massachusetts Institute of Technology, 77 Massachusetts Ave., Cambridge, MA 02139}

\author{D. Hsieh}
\email[Corresponding author. ]{dhsieh@caltech.edu}
\affiliation{Institute for Quantum Information and Matter, California Institute of Technology, Pasadena, CA 91125}
\affiliation{Department of Physics, California Institute of Technology, Pasadena, CA 91125}

\begin{abstract}
\noindent In the presence of electron-phonon coupling, an excitonic insulator harbors two degenerate ground states described by an Ising-type order parameter. Starting from a microscopic Hamiltonian, we derive the equations of motion for the Ising order parameter in the phonon coupled excitonic insulator Ta$_2$NiSe$_5$ and show that it can be controllably reversed on ultrashort timescales using appropriate laser pulse sequences. Using a combination of theory and time-resolved optical reflectivity measurements, we report evidence of such order parameter reversal in Ta$_2$NiSe$_5$ based on the anomalous behavior of its coherently excited order-parameter-coupled phonons. Our work expands the field of ultrafast order parameter control beyond spin and charge ordered materials.
\end{abstract}

\maketitle

Exploring new pathways to optically switch Ising-type electronic order parameters is a major theme of current ultrafast science. In recent years, a variety of out-of-equilibrium protocols have been developed for rapidly switching ferromagnetic \cite{FullertonScience2014, StohrNature2004, RasingNature2002}, ferrimagnetic \cite{RasingPRL2007, RasingPRL2009, KimelNature2011}, antiferromagnetic \cite{KimelNPHY2009, FiebigNPHO2016, HuberNature2019}, and ferroelectric \cite{MerlinPRL1994, RappePRL2009, CavalleriPRL2017} order parameters. However, far less is understood about the mechanisms for switching more exotic order parameters that are not of magnetic and charge dipolar type.

A particularly interesting case is the excitonic insulator (EI), a strongly correlated electronic phase realized through condensation of bound electron-hole pairs \cite{KohnPR1967}. The free energy landscapes of the complex electronic order parameter and the real lattice order parameter of an EI are typically characterized by a Mexican hat with continuous U(1) symmetry and a parabola, respectively [Fig. 1(a)]. However, strong electron-phonon coupling (EPC) induces a tilting of both the lattice and electronic potentials \cite{MSNCOMM2020, BorisPRB2017, OhtaPRB2013, SawaPRB2018, OkazakiARXIV2020}, reducing the U(1) symmetry to a discrete $Z_2$ Ising-type symmetry. Like in magnetic or charge dipole ordered ferroic materials, this leads to two degenerate ground states characterized by order parameters of equal magnitude but opposite phase [Fig. 1(b)].

There is currently no experimental method to switch nor directly measure the phase of an EI order parameter on ultrashort timescales. An alternative strategy is to measure the phase of the coupled structural order parameter. However, existing time-resolved x-ray and electron diffraction techniques are not phase sensitive. Optical phase-resolved second harmonic generation measurements \cite{CavalleriPRL2017,FiebigOL1994,FiebigAPL1995} have been used to measure the phase of structural order parameters in noncentrosymmetric ferroic materials \cite{FiebigPRL1994, FiebigNature2002, FiebigNature2007, FiebigScience2015}, but all known EI candidates are centrosymmetric \cite{AbbamonteScience2017,TakagiPRL2009}. The possible presence of 180$^{\circ}$ EI domains further complicates such measurements because domain averaging would cause overall signal cancellation.

In this Letter, we demonstrate via theory and experiment a pathway to optically switch an EI order parameter and to probe this reversal through a coherent EI order-parameter-coupled phonon (OPCP). The time evolution of the coupled EI and structural order parameters following impulsive laser excitation are derived from modeling a prototypical system Ta\textsubscript{2}NiSe\textsubscript{5}, which we assume to harbor an EI phase, with an elementary spinless two-band Hamiltonian. Our simulations reveal that the EI order parameter is stably reversed above a critical laser fluence, identifiable indirectly via a saturation of the coherent OPCP amplitude. Using time-resolved coherent phonon spectroscopy measurements, we experimentally verify this scenario and also demonstrate how switching can be controlled through the relative timing between successive laser excitation pulses.

The quasi-one-dimensional (1D) direct band-gap semiconductor Ta\textsubscript{2}NiSe\textsubscript{5} is reported to undergo an EI transition at a critical temperature $T_c =$ 328 K \cite{TakagiPRL2009,OhtaPRB2014}, accompanied by a weak orthorhombic-to-monoclinic structural distortion due to EPC \cite{OhtaPRB2013}. Impulsive laser excitation below $T_c$ has been shown to coherently excite at least five distinct Raman-active phonons with frequencies near 1, 2, 3, 3.7, and 4 THz. The 1, 2, and 4 THz modes are sensitive to the EI transition at $T_c$ \cite{KaiserSciAdv2018,MonneyPRB2018,OkazakiARXIV2020,ZWTPRL2020} and thus constitute the OPCPs, while the 3 and 3.7 THz modes are reportedly not coupled to the EI order parameter and thus serve as a control.

\begin{figure}[t]
\includegraphics[width=3.375in]{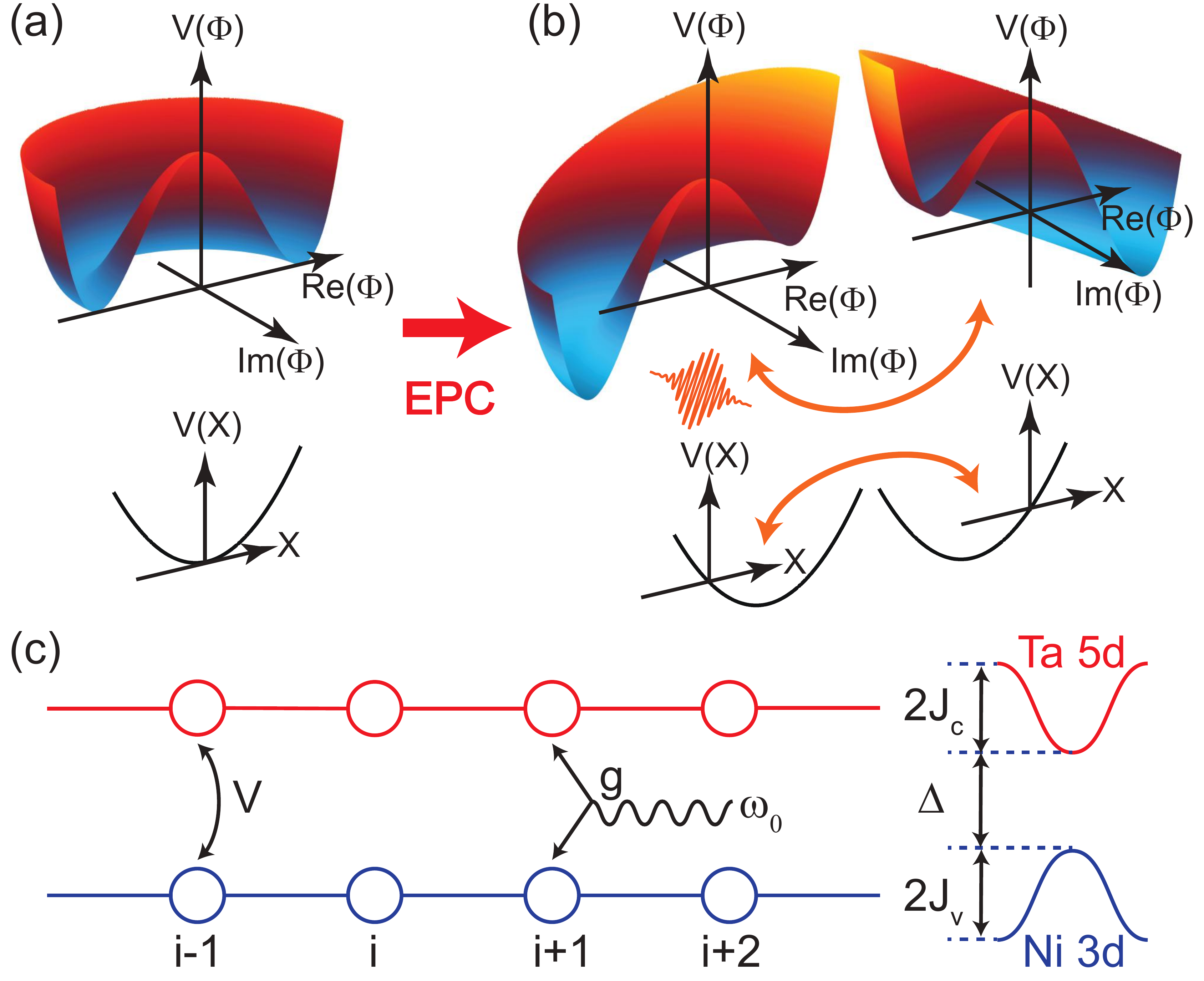}
\label{Fig1}
\caption{Schematic of the electronic and structural free energy landscapes (a) without and (b) with EPC. In the latter case, pulsed excitation can drive the system between two degenerate ground states. (c) Schematic of the 1D spinless two-band model. Red (blue) circles on each site $i$ denote Ta 5$d$ conduction band (Ni 3$d$ valence band) states. The microscopic parameters discussed in the main text are defined pictorially.}
\end{figure}

The low energy electronic structure of Ta\textsubscript{2}NiSe\textsubscript{5} consists of a conduction band with Ta 5$d$ orbital character and a valence band with Ni 3$d$-Se 4$p$ hybridized orbital character. The EI instability is well captured by a 1D spinless two-band Hamiltonian with EPC \cite{ZenkerPRB2014, TanakaPRB2018, EcksteinPRL2017, OhtaPRB2018} [Fig. 1(c)],
\begin{equation}\label{eqn:1}
\begin{split}
H = \sum_{k}(\epsilon_{k}{c_k}^{\dagger}c_k+\mu_{k}{v_k}^{\dagger}v_k) + \sum_{i}[V{c_i}^{\dagger}c_i{v_i}^{\dagger}v_i\\
+{\omega}_0{b_i}^{\dagger}b_i+g({b_i}^{\dagger}+b_i)({c_i}^{\dagger}v_i+{v_i}^{\dagger}c_i)],
\end{split}
\end{equation}

\noindent where ${c_k}^{\dagger},c_k$ and ${v_k}^{\dagger},v_k$ are the fermionic creation and annihilation operators for conduction and valence band electrons with momentum $k$, respectively, and ${b_i}^{\dagger},b_i$ are the bosonic creation and annihilation operators for an OPCP mode of energy ${\omega}_0$ at site $i$. The conduction and valence band dispersions are given by ${\epsilon}_k=\frac{\Delta}{2}+2J_c{\sin{(ka/2)}}^2$ and ${\mu}_k=-\frac{\Delta}{2}-2J_v{\sin{(ka/2)}}^2$, respectively, where $2J_c$ and $2J_v$ are their bandwidths and $\Delta$ is the band gap \cite{EcksteinPRL2017,OhtaPRB2018}. An on-site interband electron-electron interaction term $V$ drives the excitonic pairing, and an EPC term $g$ couples the electronic and phononic subsystems. These microscopic parameters have been experimentally determined \cite{EcksteinPRL2017,OhtaPRB2018}.

From Eq.(\ref{eqn:1}), we derive the equations of motion for the EI and structural order parameters, defined as ${\Phi}_i={\langle}{c_i}^{\dagger}v_i{\rangle}$ and $X_i={\langle}{b_i}^{\dagger}+b_i{\rangle}$, respectively, in two steps. We first derive the exact expression for the nonequilibrium free energy functional of $\Phi$ and $X$ in the Keldysh path integral framework, where we include the light excitation via Peierls substitution. Then we obtain the equations of motion as the saddle point of the free energy, ignoring population in the conduction bands, spatial fluctuations of the order parameter field, and higher-order contributions $O(\Phi^6)$ (see Supplemental Material \cite{SM}),
\begin{equation} \label{eqn:2}
i{\partial}_t{\Phi}=[-D(\nabla-iqA)^2+m+U{|{\Phi}|}^2]{\Phi}+2g'X,
\end{equation}
\begin{equation} \label{eqn:3}
{{\partial}_t}^2{X}=-({{\omega}_0}^2+\frac{2g^2{\omega}_0}{V})X-2g{\omega}_0\operatorname{Re}(\Phi).
\end{equation}
\noindent Here $D$ is an effective diffusion coefficient, $q$ is the electron charge, $A$ is the light vector potential, $g'$ is the renormalized EPC coefficient for the electronic channel, and $m$ and $U$ are the second- and fourth-order expansion coefficients of the electron-electron interaction term, respectively. These parameters are all functions of $J_c$, $J_v$, $\Delta$, $g$, and $V$ \cite{SM}. We further introduce two phenomenological constants $\gamma_e$ and $\gamma_{ph}$ to account for damping of electronic and structural modes due to thermal fluctuation, whose values can be experimentally determined \cite{SM}.

\begin{figure*}[t]
\includegraphics[width=6.75in]{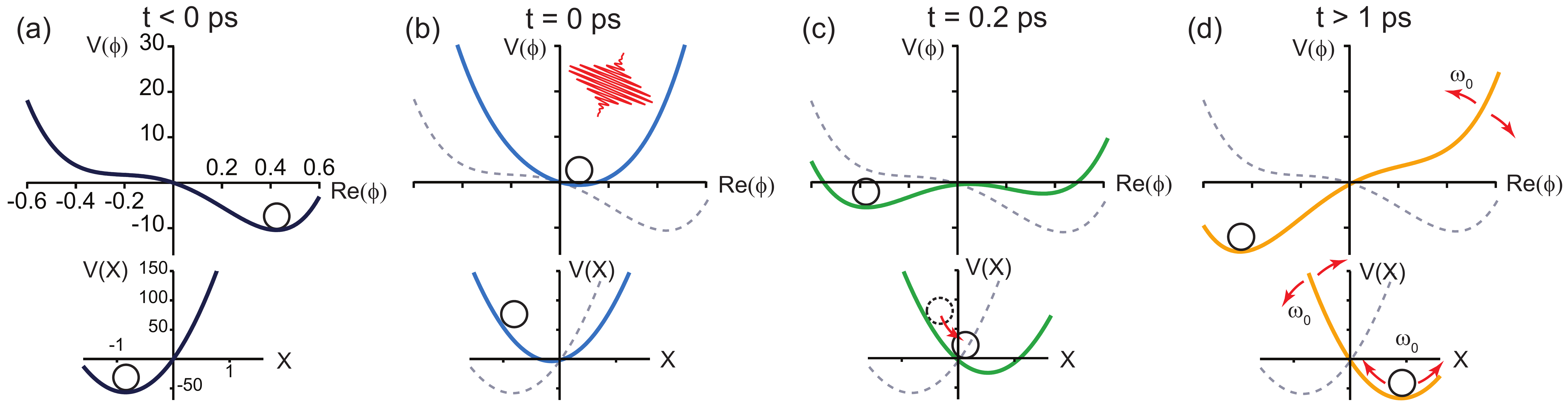}
\label{Fig2}
\caption{Simulation results of $V(\Phi)$ and $V(X)$ (defined in main text) for $F > F_c$ with experimentally determined parameters \cite{SM}. Snapshots of the potential landscapes (solid lines) and the electronic and structural order parameters (circles) are shown (a) in the equilibrium state (reproduced as dashed lines in (b)-(d)), (b) at the moment of excitation, (c) during transit into the reversed state, and (d) in the reversed state before equilibration, where both potentials are modulated at the phonon frequency (red arrows). Axes' scales are the same in all panels.}
\end{figure*}

We first present a qualitative picture of how order parameter reversal occurs in our model. In equilibrium the electronic potential $V(\Phi)=\frac{1}{2}m{\Phi}^2+\frac{1}{4}U{\Phi}^4+2g'X{\Phi}$ has either a tilted parabolic ($m >$ 0) or tilted Mexican-hat ($m <$ 0) form capturing the absence or presence of exciton condensation. For homogeneous optical excitation, one can ignore spatial derivatives of $\Phi$. For pulsed excitation with infrared light, whose frequency well exceeds the electronic Higgs-Goldstone \cite{EcksteinPRL2017,WernerPRB2020} and OPCP mode frequencies, one can also average out the fast oscillations of the perturbation and retain only its Gaussian envelope. Under these conditions one can make the simplification $m(t)=-D(\nabla-iqA)^2 + m$ ${\triangleq}f(t)+m$, where $f(t)={\alpha}F\exp{(-\frac{4\ln({2})t^2}{\sigma^2})}$. Here $\sigma$ is the temporal width of the Gaussian pulse, $\alpha$ is a positive constant scaling factor that can be calculated analytically \cite{SM}, and $F$ is the pump fluence, the only tunable parameter in our model. In the EI phase ($m <$ 0), optical excitation therefore acts to instantaneously increase $m(t)$. The subsequent reduction of the EI order parameter, which occurs on a timescale much shorter than $2\pi/\omega_0$, results in a sudden shift in the lattice potential $V(X)=\frac{1}{2}{\omega_0}^2{X}^2+2g{\omega_0}\operatorname{Re}(\Phi)X$ due to EPC \cite{SM}, launching coherent oscillations through displacive excitation. In the low fluence regime, where the phonon oscillation amplitude is small enough such that $X$ does not change sign, the direction of tilt of both $V(\Phi)$ and $V(X)$ remains unchanged and so no switching occurs. However, above a critical fluence $F_c$, where the phonon oscillation amplitude becomes large enough to change the sign of $X$, the tilting of both potentials is reversed and the system can relax into the switched state.

Numerical simulations of our model using experimentally determined material parameters for Ta\textsubscript{2}NiSe\textsubscript{5} and $\sigma$ = 100 fs were carried out in the EI phase \cite{SM}. Figure 2 displays simulation results for $F$ slightly greater than $F_c$. At the moment of excitation $t = 0$ [Fig. 2(b)], there is an instantaneous change in $V(\Phi)$ from Mexican-hat to parabolic form caused by the light-induced enhancement of $m(t)$. The EI order parameter evolves rapidly to the new potential minimum with overdamped dynamics and is quenched within the pulse duration. This leads to a rapid shift in $V(X)$, launching coherent oscillations of the underdamped OPCP, which shakes the electronic potential via EPC at the phonon frequency. Once the pulsed excitation is over, $V(\Phi)$ recovers a Mexican-hat form. However, as $X$ crosses zero within the first half period of oscillation [Fig. 2(c)], the tilting of $V(\Phi)$ is reversed with respect to the pre-pumped ($t <$ 0) case [Fig. 2(a)], sending $\Phi$ toward the new minimum on the negative side. As $\Phi$ crosses zero, the tilt of $V(X)$ is also reversed due to EPC, thus pushing $X$ to the new minimum on the positive side. The system then continues to oscillate about the reversed minima at the phonon frequency until the OPCP is damped out [Fig. 2(d)]. Note that this model treats light as a coherent drive without considering heating- and cooling-induced changes in $m(t)$. However, accounting for the latter merely shifts $F_c$ \cite{SM}.

In lieu of probing the phase of $X$ and $\Phi$, we propose that the reversal can be identified via the pump fluence dependence of the OPCP amplitude. We solve Eqs. (\ref{eqn:2}) and (\ref{eqn:3}) to obtain the time evolution of $\Phi$ and $X$ and then define the OPCP amplitude by its peak height in the fast Fourier transform (FFT) of $X$. A conventional Raman-active phonon is coherently launched through either displacive excitation or impulsive stimulated Raman scattering (ISRS) \cite{MerlinPRB2002,MisochkoPLA2011} with an amplitude that is linearly proportional to $F$. For $F < F_c$, the amplitude of an OPCP also scales linearly with $F$. In this regime the structural order parameter simply oscillates about the initial potential minimum and thus behaves like a conventional phonon. But once the initial displacement of $V(X)$ is large enough to enable escape to the opposite minimum ($F > F_c$), the amplitude ceases to grow. Order parameter reversal is thus marked by a saturation in the amplitude versus fluence curve.

We argue that experimental evidence for this phenomenon already exists in published studies of Ta$_2$NiSe$_5$. Werdehausen $et$ $al.$ \cite{KaiserSciAdv2018} performed ultrafast optical reflectivity measurements at $T$ = 120 K using a pump photon energy of 1.55 eV and observed clear coherent oscillations of the 1 THz OPCP as well as the uncoupled 3 THz phonon (ISRS). The pump fluence dependence of these two mode amplitudes is reproduced in Fig. 3. The 3 THz mode scales linearly with $F$, consistent with its assignment as a conventional phonon. In contrast, the 1 THz mode scales linearly with $F$ only at low fluences and then saturates above $\sim$0.4 mJ/cm$^2$, consistent with an order parameter reversal. By overlaying our simulation results \cite{SM} atop these experimental curves, we find close agreement (Fig. 3).

Our theory also predicts that the exciton condensate should be transiently quenched [$m(t) \rightarrow 0$] above a critical fluence $F = F^*$ where the condition $f(t) = |m|$ is satisfied. Eqs. (\ref{eqn:2}) and (\ref{eqn:3}) do not constrain $F^*$ to coincide with $F_c$ and our simulation shows that $F^*$ is clearly lower than $F_c$ in Ta$_2$NiSe$_5$ (Fig. 3). Recently Tang $et$ $al.$ \cite{ZWTPRL2020} performed time- and angle-resolved photoemission spectroscopy (tr-ARPES) measurements on Ta$_2$NiSe$_5$ at $T$ = 30 K and tracked the dynamics of the charge gap, a measure of $\Phi$, immediately after pumping with 1.77 eV light polarized perpendicular to the chain direction (equivalent geometry to Ref. \cite{KaiserSciAdv2018}). They found that the instantaneous gap size decreases linearly with increasing pump fluence and saturates above 0.29 mJ/cm$^2$, which was interpreted as the point where $\Phi$ transiently collapses. The fact that this fluence is lower than 0.4 mJ/cm$^2$, and is expected to be even lower if the experiment were conducted at 120 K, is consistent with our theory.

\begin{figure}[t]
\centering
\includegraphics[width=3.375in]{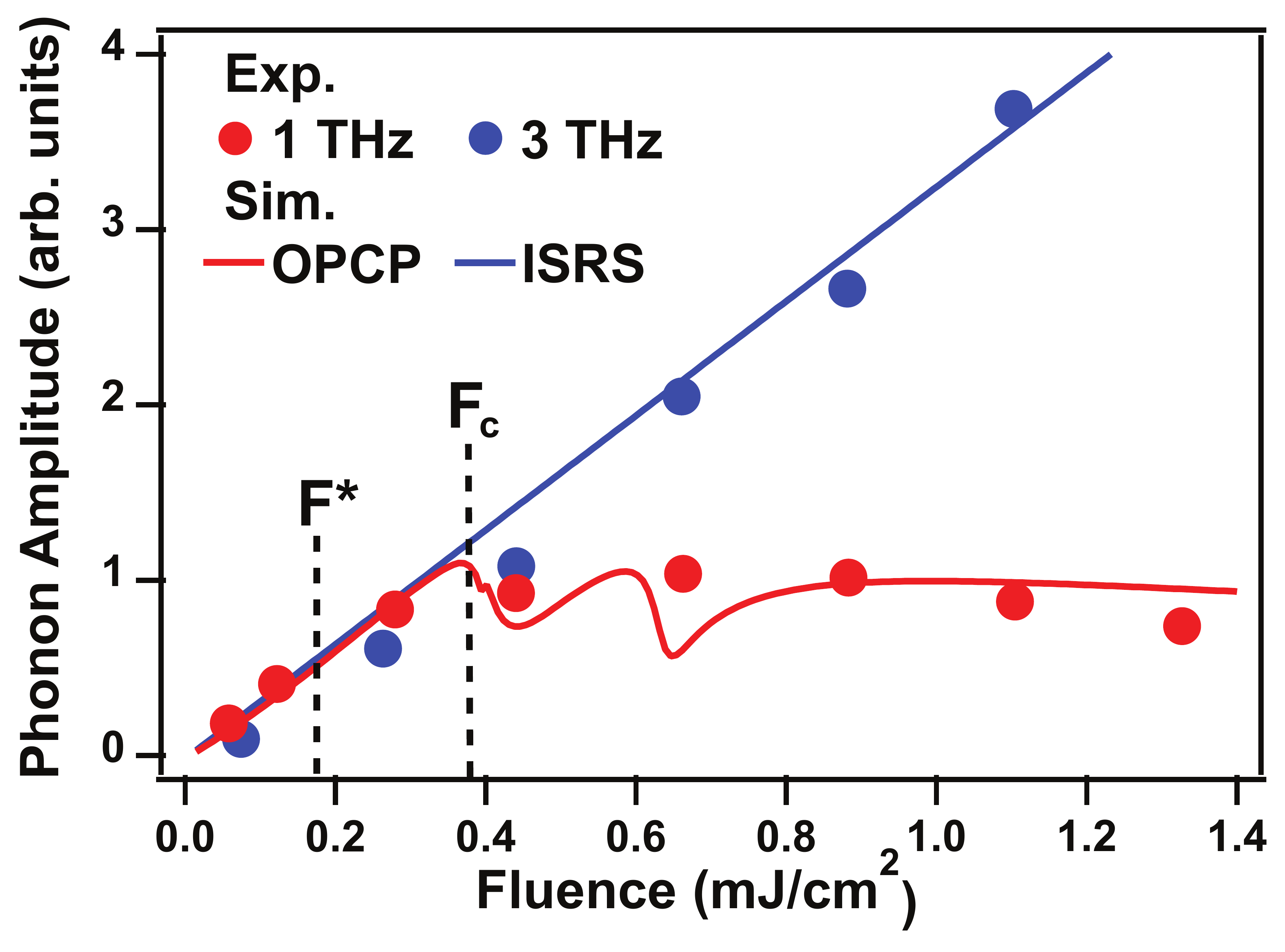}
\label{Fig3}
\caption{Experimental pump fluence dependence of the 1 THz (red circles) and 3 THz (blue circles) coherent phonon amplitudes in Ta$_2$NiSe$_5$ reproduced from Ref.\cite{KaiserSciAdv2018}. Simulation results for an OPCP (red line) and a conventional ISRS phonon (blue line) are overlaid and horizontally scaled ($\alpha \approx 700$) to match the experimental data. Vertical dashed lines mark the calculated $F_c$ and $F^*$. The non-monotonic behavior of the OPCP amplitude just above $F_c$ arises from strong feedback between $\Phi$ and $X$ immediately after excitation \cite{SM}.}
\end{figure}

The dynamical nature of the order parameter reversal process suggests that it can be controlled not merely by the total pump energy deposited, but also by its distribution in time. To show this, we consider a situation where the sample is pumped by two identical pulses separated by time $\delta t$, with individual fluences $F < F_c$ but $2F > F_c$. For $\delta t \rightarrow 0$ the system is effectively pumped by a single pulse exceeding $F_c$ and so reversal occurs, while for $\delta t \rightarrow \infty$, the system relaxes back to the initial ground state before the second pulse arrives and so no reversal occurs. To qualitatively understand the behavior at intermediate $\delta t$ values where the system is still dynamically evolving when the second pulse arrives, we recall that in the single pulse case, switching occurs once the OPCP amplitude is large enough to change the sign of $X$. Therefore, in the two-pulse case, switching possibly occurs if coherent oscillations of the OPCP induced by the first pulse can be sufficiently amplified by the second time-delayed pulse.

\begin{figure}[t]
\includegraphics[width=3.375in]{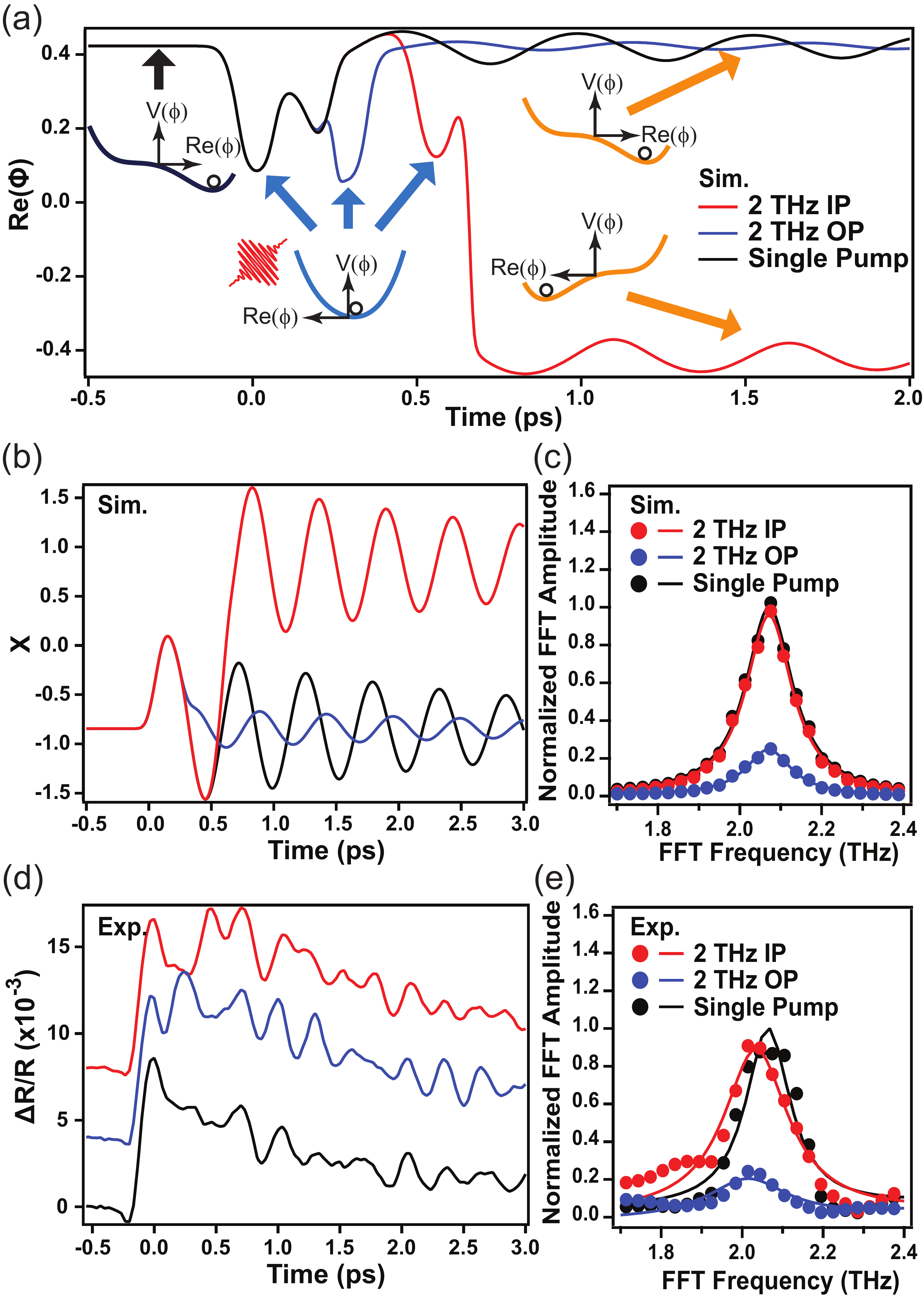}
\label{Fig4}
\caption{Simulated time evolution of (a) Re($\Phi$) and (b) $X$ following single-pulse pumping (black) and two-pulse OP (blue) and IP (red) pumping of the 2 THz phonon using $F = 0.96 F_c$ and the same microscopic parameters as in Figs. 2 and 3. The calculated instantaneous electronic potential is displayed at several select times. (c) Normalized FFT of the traces shown in (b). Each curve is normalized by the peak value of the single-pulse pumping curve. (d) Reflectivity transients measured from Ta$_2$NiSe$_5$ under the same pumping conditions used in the simulations. Curves are vertically offset for clarity. (e) Normalized FFT of the traces shown in (d).}
\end{figure}

Previous studies have shown that impulsively and displacively excited conventional Raman-active phonons can be coherently amplified (suppressed) by a second pump pulse when $\delta t$ is an integer (half-integer) multiple of the phonon period \cite{HaseAPL1996, HaseJJAP1998, Reis2005, MisochkoJPCM2007, Wu2007, Cheng2017}, dubbed in-phase (IP) and out-of-phase (OP) pumping respectively. Therefore we simulated the effects of both IP and OP pumping on the order parameters of Ta$_2$NiSe$_5$ using the same material parameters as before. We chose to simulate the OPCP at $\omega_0/2\pi$ = 2 THz rather than at 1 THz because recent tr-ARPES data show that the most pronounced modulations of the valence band maximum occur at 2 THz, suggesting strong coupling to $\Phi$ \cite{OkazakiARXIV2020}. As shown in Fig. 4(a), pumping by a single pulse with $F$ slightly less than $F_c$ causes a rapid but incomplete reduction of $\Phi$, followed by a slower recovery back to its original value on a timescale set by the damping of the 2 THz phonon. For the two-pulse case, OP pumping of the 2 THz phonon similarly leads to partial suppression of $\Phi$ without reversal, but reversal is achieved with IP pumping. This phenomenon is again manifested through an unconventional behavior of the OPCP. As shown in Figs. 4(b) and (c), OP pumping leads to suppression of the 2 THz phonon amplitude relative to the single pump case, resembling a conventional phonon because the oscillation is around the initial potential minimum. But, in contrast to conventional behavior, IP pumping does not lead to further amplification once $X$ is excited to the opposite minimum.

To verify this prediction, we performed transient optical reflectivity measurements on Ta$_2$NiSe$_5$ single crystals \cite{SM} using two identical pump pulses ($\sigma$ = 80 fs) with variable $\delta t$. The light was polarized perpendicular to the chain direction and the fluence of each pulse was tuned slightly below $F_c$ to match our simulations. We chose a pump photon energy of 1 eV to enhance the 2 THz oscillations \cite{SM}. Figure 4(d) shows the fractional change in reflectivity ($\Delta R/R$) versus time for both IP and OP pumping of the 2 THz phonon, as well as for pumping with only a single pulse. All three curves exhibit fast ($\sim$1 ps) exponential decay following pump excitation, corresponding to the charge relaxation process. Oscillations from the beating of several coherently excited phonons are also clear. A FFT of the data shows the most pronounced peaks at 2, 3, and 3.8 THz \cite{SM}. A focus on the 2 THz mode reveals that OP pumping strongly suppresses its amplitude relative to the single-pump case whereas IP pumping does not amplify it [Fig. 4(e)], in quantitative agreement with our simulations [Fig. 4(c)]. In contrast, strong amplification occurs for the 3 and 3.8 THz modes \cite{SM}, consistent with their uncoupled nature.

Our field theory description of the EI order parameter goes beyond the phenomenological time-dependent Landau theory \cite{JohnsonPRL2014, GedikPRL2019, MihalovicNPHY2010, HsiehPRL2018} in that it allows the order parameter to explore the tilted Mexican-hat potential in the complex plane and can be naturally linked to microscopic parameters of the underlying lattice model. While more details including extension beyond the mean-field limit ($\nabla\neq0$), temperature dependence with $T>0$, diffusion perpendicular to the surface \cite{MihalovicNPHY2010}, time-dependent damping \cite{JohnsonPRL2014}, and anharmonic phonon coupling \cite{HasePRL2002} can be added to refine the simulations, our minimal microscopic theory already captures the most salient physics and experimental features. These ideas and dynamical protocols apply not only to excitonic insulators, but also to any system featuring a continuous-symmetry-breaking electronic order parameter induced by coupling to a structural order parameter, such as a charge ordered system coupled to a Peierls distortion or an orbital ordered system coupled to a Jahn-Teller distortion. Therefore the OPCP behavior revealed here may be a general fingerprint of electronic order parameter switching.

\begin{acknowledgements}
We thank Rick Averitt, Edoardo Baldini, Swati Chaudhary, Nuh Gedik, Xinwei Li, Tianwei Tang and Alfred Zong for useful discussions. D.H. and J.G.C. acknowledge support from the DARPA DRINQS program (Grant No. D18AC00014). D.H. also acknowledges support for instrumentation from the Institute for Quantum Information and Matter, a NSF Physics Frontiers Center (PHY-1733907). M.B. acknowledges the support from the Department of Energy under Award No. DE-SC0019166. T.K. acknowledges the support by the Yamada Science Foundation Fellowship for Research Abroad and JSPS Overseas Research Fellowships.
\end{acknowledgements}


%

\end{document}


\title{Supplemental Material for ``Signatures of ultrafast reversal of excitonic order in Ta\textsubscript{2}NiSe\textsubscript{5}"}

\author{H. Ning}
\affiliation{Institute for Quantum Information and Matter, California Institute of Technology, Pasadena, CA 91125}
\affiliation{Department of Physics, California Institute of Technology, Pasadena, CA 91125}

\author{O. Mehio}
\affiliation{Institute for Quantum Information and Matter, California Institute of Technology, Pasadena, CA 91125}
\affiliation{Department of Physics, California Institute of Technology, Pasadena, CA 91125}

\author{M. Buchhold}
\affiliation{Institute for Quantum Information and Matter, California Institute of Technology, Pasadena, CA 91125}
\affiliation{Department of Physics, California Institute of Technology, Pasadena, CA 91125}

\author{T.Kurumaji}
\affiliation{Department of Physics, Massachusetts Institute of Technology, 77 Massachusetts Ave., Cambridge, MA 02139}

\author{G. Refael}
\affiliation{Institute for Quantum Information and Matter, California Institute of Technology, Pasadena, CA 91125}
\affiliation{Department of Physics, California Institute of Technology, Pasadena, CA 91125}

\author{J. G. Checkelsky}
\affiliation{Department of Physics, Massachusetts Institute of Technology, 77 Massachusetts Ave., Cambridge, MA 02139}

\author{D. Hsieh}
\affiliation{Institute for Quantum Information and Matter, California Institute of Technology, Pasadena, CA 91125}
\affiliation{Department of Physics, California Institute of Technology, Pasadena, CA 91125}

\maketitle

\tableofcontents

\section{I. Derivation of the microscopic dynamical equations}
In order to connect the dynamics of the OPCP and the exciton condensate with the microscopic description of Ta$_2$NiSe$_5$, we start from a commonly used two-band semiconductor Hamiltonian with spinless fermions \cite{ZenkerPRB2014, TanakaPRB2018, EcksteinPRL2017, OhtaPRB2018} and inter-band interactions,
\begin{equation}
\label{Eqn3}
\begin{split}
H = \sum_{k}(\epsilon_{k}{c_k}^{\dagger}c_k+\mu_{k}{v_k}^{\dagger}v_k) + \sum_{i}(V{c_i}^{\dagger}c_i{v_i}^{\dagger}v_i\\
+{\omega}_0{b_i}^{\dagger}b_i+g({b_i}^{\dagger}+b_i)({c_i}^{\dagger}v_i+{v_i}^{\dagger}c_i)) 
\end{split}
\end{equation}
The bands are formed by the quasi-one-dimensional lattice and $i, k$ are the corresponding lattice sites and momentum. Each operator and parameter is defined in the main text. We also set $\hbar=1$ for simplicity.

The equations of motion for the complex exciton order parameter $\Phi_i={\langle}c^\dagger_i v_i{\rangle}$ and the real lattice displacement $X_i={\langle}b^\dagger_i+b_i{\rangle}$ can be obtain by expressing Eq.~\eqref{Eqn3} in terms of a nonequilibrium path integral (i.e. in the Keldysh framework) and introducing $\Phi$ as a dynamic, bosonic Hubbard-Stratonovich field. This allows one to integrate out the fermionic modes and derive the equations of motion for $\Phi, X$ via saddle point equations. 

In a path integral approach, the partition function $Z$ corresponding to the Hamiltonian in Eq.~\eqref{Eqn3} is formally obtained from a field integral of the form
\begin{align}
Z=\int\mathcal{D}[\{\bar c_i, c_i, \bar v_i, v_i, b^*_i, b_i, \Phi^*_i, \Phi_i\}]e^{iS},
\end{align}
where $\mathcal{D}$ represents the common field integral measure and $\bar c_i, c_i, \bar v_i, v_i$ are independent Grassmann fields, representing the fermion modes in conduction and valence bands, and $b^*_i, b_i, \Phi^*_i, \Phi_i$ are complex fields, corresponding to the phonon and exciton condensate modes. The Keldysh action $S$ is obtained in the canonical way \cite{KamenevBook2011} and reads as $S=S_{\text{f}}+S_{\text{b}}$ with the fermion part 
\begin{align}
S_{\text{f}}=\sum_{l,m}\int_t \bar\Psi_{l,t}\left(\begin{array}{cc}C^{-1}_{l,m,t}& M_{l,t}\delta_{l,m}\\ M^*_{l,t}\delta_{l,m}& W^{-1}_{l,m,t}\end{array}\right)\Psi_{m,t}^T
\end{align}
and the boson part 
\begin{align}
S_{\text{b}}=&\sum_l\int_t\Big[(b^*_{c,l,t}, b^*_{q,l,t})B^{-1}_{l,t} \left(\begin{array}{c}b_{c,l,t}\\ b_{q,l,t}\end{array}\right)\\&-V \left(\Phi^*_{c,l,t}\Phi_{q,l,t}+\Phi^*_{q,l,t}\Phi_{c,l,t}\right)\Big].\nonumber
\end{align}
Here, $\Psi_{l,t}=(c_{1,l,t},c_{2,l,t},v_{1,l,t},v_{2,l,t})$ is the fermion spinor in Keldysh space and each field carries an index triplet $(i,l,t)$, which labels Keldysh component $i$ ($i=1,2$ for Grassmann fields and $i=c,q$ for complex fields), lattice site $l$ and time $t$. The matrices $B,C,W$ are the Keldysh space Green's function for phonons, conduction band, and valence band. They are diagonal in frequency and momentum space 
\begin{align}
C^{-1}_{k,\omega}=\left(\begin{array}{cc} 0 &\omega-\epsilon_k-i\eta\\ 
\omega-\epsilon_k+i\eta& 2i\eta \tanh(\omega/2T)\end{array}\right) 
\end{align}
and $W^{-1}_{k,\omega}$ is identical with $\epsilon_k\rightarrow\mu_k$. Also, 
\begin{align}
B^{-1}_{k,\omega}=\left(\begin{array}{cc} 0 &\omega-\omega_0-i\eta\\ 
\omega-\omega_0+i\eta& 2i\eta \tanh(\omega/2T)\end{array}\right)
\end{align}
with $\eta\rightarrow0^+$. The matrix $M_{l,t}$ describes the local coupling of the fermions to the exciton and phonon fields 
\begin{align}
M_{l,t}=\left[\mathbf{1}\big(V\Phi_{c,l,t}+gX_{c,l,t}\big)+\sigma_x \big(V\Phi_{q,l,t}+gX_{q,l,t}\big)\right].
\end{align}

The fermion part $S_{\text{f}}$ is quadratic in Grassmann fields and integration according to Grassmann calculus yields the formal expression
\begin{align}\label{Eq10}
S_{\text{f}}=-i \text{Tr}\left(\log\left[\mathbf{1}-CMWM^*\right]\right)=i\sum_{n=1}^\infty\frac{1}{n}\text{Tr}\left(CMWM^*\right)^n,
\end{align}
where the trace includes the sum over Keldysh indices and lattice sites and an integral over time. In order to eliminate the phonon field from the nonlinear action $S_{\text{f}}$, one performs a polaron-type shift $\Phi_{c/q,l,t}\rightarrow \Phi_{c/q,l,t}-\frac{g}{V}X_{c/q,l,t}$. For small exciton field amplitudes $\Phi$, Eq.~\eqref{Eq10} can be expanded up to fourth order in the fields ($n\le2$) and in powers of derivatives. The equations of motion for the exciton condensate and the displacement, which are represented by the classical fields $\Phi_{c,l,t}, X_{c,l,t}$ are obtained via the saddle-point equations (and their complex conjugates)
\begin{align}
\frac{\delta S}{\delta b_{q,l,t}}=\frac{\delta S}{\delta \Phi_{q,l,t}}=0.
\end{align}

This yields the equations of motion by further introducing the perturbation of light via Peierls substitution, which is argued to work better for an electronically localized system \cite{OhtaPRB2018}, and assuming the lattice constant $d$ = 1:

\begin{align} 
iZ{\partial}_t{\Phi}&=(-\tilde D(\nabla-iqA)^2+\tilde m+\tilde U{|{\Phi}|}^2){\Phi}+\frac{2g}{V}X,\\
{{\partial}_t}^2{X}&=-({{\omega}_0}^2+\frac{2g^2{\omega}_0}{V})X-2g{\omega}_0\operatorname{Re}(\Phi),
\end{align}
with the parameters $\tilde D, \tilde m, \tilde U, Z$ depending on integrals over Green's functions and therefore on the band structure of the material and the temperature $T$ of the system. Assuming a band gap $\Delta$ and $k_BT<\Delta$, i.e. negligible population in the conduction bands, we can write out these parameters
\begin{align}
\tilde m&=1-\frac{2V}{\sqrt{\Delta(2J_c+2J_v+\Delta)}},\\
\tilde D&=\frac{2J_cJ_vV}{{\sqrt{\Delta(2J_c+2J_v+\Delta)}}^3},\\
Z&=V\left(\frac{2J_c+2J_v+2\Delta}{{\sqrt{\Delta(2J_c+2J_v+\Delta)}}^3}\right),\\
\tilde U&=\left(\frac{2(3(J_c+J_v)^2+4\Delta(J_c+J_v)+2{\Delta}^2)}{{\sqrt{\Delta(2J_c+2J_v+\Delta)}}^5}\right)V^3.
\end{align}
The equations in the main text are obtained via $m=\tilde m/Z$, $U=\tilde U/Z$, $D=\tilde D/Z$, $g'=g/(ZV)$, and $f=\tilde Dq^2A^2/Z{\triangleq} {\alpha}F\exp{(-\frac{4\ln{2}t^2}{\sigma^2})}$.

Also, we can obtain real valued equations of motion by defining $\Phi=\phi+i\eta$. To characterize the dephasing of the phononic and electronic channels, we also add phenomenological decaying terms to both branches. It is straightforward to add a $-2{\gamma}_{ph}{\partial}_tX$ term to the structural dynamical equation and ${\gamma}_{ph}$ can be determined experimentally. For the complex electronic order parameter, on the other hand, we rewrite the order parameter dynamical equation as followed:
 \begin{equation}
i{\partial}_t{\Phi}=((-D(\nabla-iqA)^2+m+U{|{\Phi}|}^2){\Phi}+2g'X)(1-i\gamma_e),
\end{equation}
Here $\gamma_e$ is dimensionless. It expresses the ratio of dephasing dynamics due to a non-zero temperature to the coherent dynamics of the order parameter. Both dynamics are generated by the same effective free energy functional. It determines the dephasing time of the electronic Higgs/Goldstone oscillations as validated in Section II. 

Also, we ignore the spatial diffusion and the phonon frequency shift due to an order of magnitude estimate $g\approx{\omega}_0{\ll}V$. With all of the above rectifications we have:
\begin{equation}
{\partial}_t{\phi}=(f+m+U({{\phi}}^2+{{\eta}}^2)){\eta}-{\gamma}_e((f+m+U({{\phi}}^2+{{\eta}}^2)){\phi}+2{g'}X),
\end{equation}
\begin{equation}
{\partial}_t{\eta}=-(f+m+U({{\phi}}^2+{{\eta}}^2)){\phi}-2{g'}X-{\gamma}_e(f+m+U({{\phi}}^2+{{\eta}}^2)){\eta},
\end{equation}
\begin{equation}
{{\partial}_t}^2{X}=-{{\omega}_0}^2X-2g{\omega}_0\phi-2{\gamma}_{ph}{\partial}_tX.
\end{equation}

We then construct the initial conditions for the above equations, which guarantee that $X$ is static and $\Phi$ remains real and static before the light excitation:
\begin{equation}\label{eqn:20}
\phi|_{t=0}=\sqrt{\frac{4gg'/{{\omega}_0}-m}{U}}. 
\end{equation}
\begin{equation}\label{eqn:17}
{\eta}|_{t=0}=0.
\end{equation}
\begin{equation}\label{eqn:15}
\partial_tX|_{t=0}=0,
\end{equation} 
\begin{equation}\label{eqn:19}
X|_{t=0}=-\frac{2g}{{\omega}_0}\sqrt{\frac{4gg'/{{\omega}_0}-m}{U}},
\end{equation}

\begin{figure*}[t]
\includegraphics[width=6.75in]{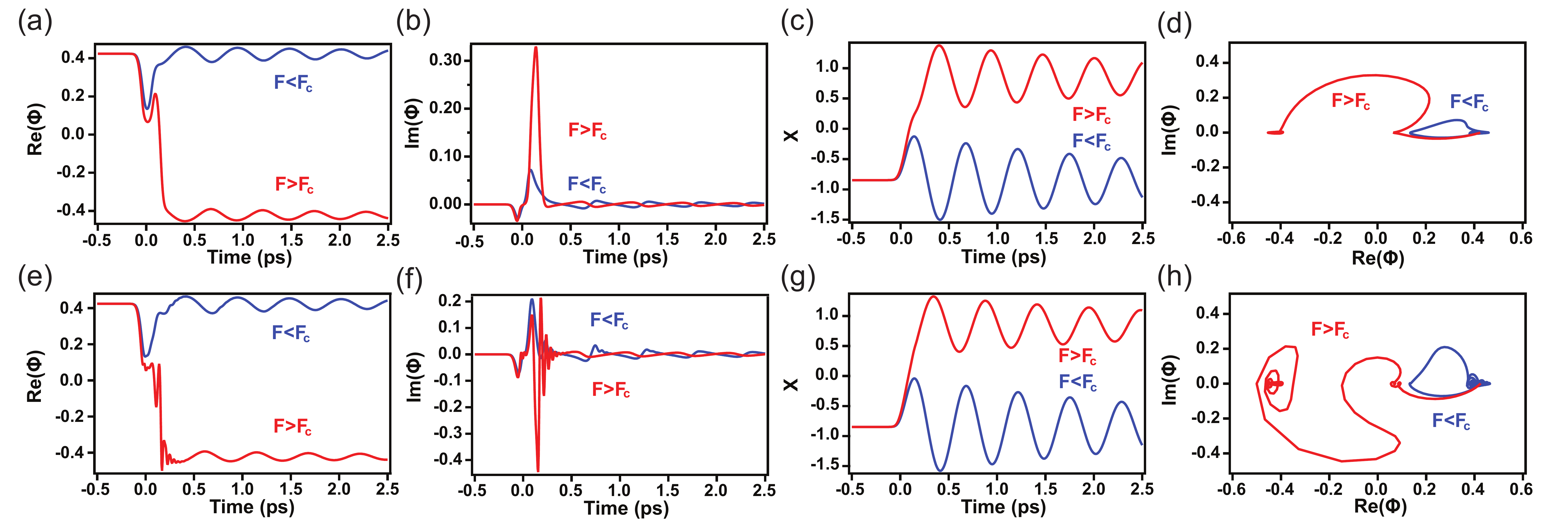}
\label{FigS1}
\caption{Simulated time evolution of the electronic and lattice order parameters in the (a)-(d) overdamped ($\gamma_e$=1) and (e)-(h) underdamped ($\gamma_e$=0.1) cases. In each case, data are shown with fluences below and above the critical fluence, and the model parameters are set to ${\omega}_0/(2\pi)=2$ THz, $g=2$ THz, $V=60$ THz, $\Delta=J_c=J_v=40$ THz, and ${\gamma}_{ph}=0.3$ THz. (a),(e) Time evolution of the real part of the electronic order parameter. (b),(f) Time evolution of the imaginary part of the electronic order parameter. (c),(g) Time evolution of the lattice order parameter. (d),(h) Trajectory of the electronic order parameter.}
\end{figure*}

After establishing the equations and the initial conditions, we can numerically solve the differential equations and trace the dynamics of the complex electronic order parameter $\Phi$ and the real structural order parameter $X$ [Fig. S1], as well as the dynamical free energy landscapes. By taking the fast Fourier transform (FFT) of $X$ in the time interval from 0 ps to 20 ps with different pumping fluence values, we obtain the order-parameter-coupled phonon (OPCP) amplitude versus fluence curves [Fig. S2]. 

To simulate the two-pulse pumping situation, one simply adds another $f$ term to Eq. (17) and (18) that is identical to the first, except that this $f$ term is centered at the time when the second pulse arrives at the sample. The initial conditions are the same. Here we apply a FFT to $X$ in the time interval between the arrival of the second pulse and 20 ps thereafter. We thus obtain the OPCP amplitude versus fluence at different time delays [Fig. S5].

\section{II. Determination of the microscopic parameters}
From our experiment we get ${\omega}_0/(2\pi)=2$ THz. The chosen microscopic parameter values in Ref.\cite{EcksteinPRL2017, OhtaPRB2018} reproduce the equilibrium band structure qualitatively well, therefore we adopt these and set $g=2$ THz, $V=60$ THz, $\Delta=40$ THz, $J_c=J_v=40$ THz, $m=-17$ THz, $U=132$ THz, $D=13.3$ THz, and $g'=0.83g$. The Higgs mode frequency $-2m=34$ THz qualitatively matches the gap size $\Delta$ \cite{KaiserSciAdv2018}. The corresponding fast oscillation is beyond the time resolution of our experimental setup, and hence cannot be resolved. The phonon dephasing time we measured is approximately 3 ps, thus ${\gamma}_{ph}\approx0.3$ THz. This set of parameter choices is self-consistent but may not be unique. We also demonstrate that the specific choice of the above microscopic parameters does not change the main conclusion of this paper, i.e. the reversal of the EI order [Fig. S2(b)]. We simulate our incident light as a $\sigma=100$ fs Gaussian irradiating the sample at $t=0$. The pump fluence $F$ is thus the only tunable parameter. 

\begin{figure*}[t]
\centering
\includegraphics[width=6.75in]{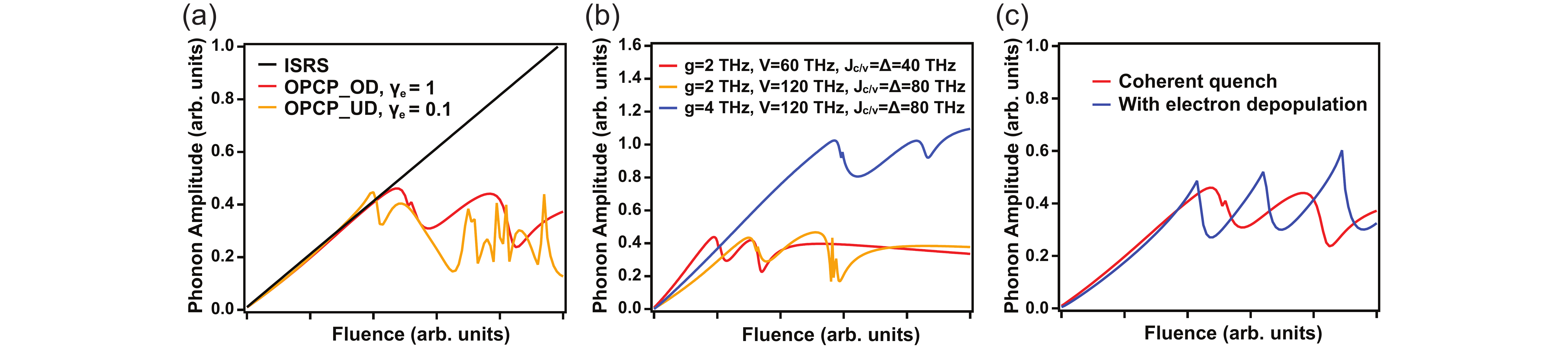}
\label{FigS2}
\caption{Fluence dependence of the OPCP with various choices of the microscopic model parameters. (a) OPCP amplitude versus fluence with different electronic decay times. The red curve corresponds to the overdamped case ($\gamma_e=1$) and the yellow curve corresponds to the underdamped case ($\gamma_e=0.1$) with $g=2$ THz, $V=60$ THz, $\Delta=J_c=J_v=40$ THz. The black curve characterizes the ISRS/DECP amplitude versus fluence. (b) OPCP amplitude versus fluence in the overdamped case ($\gamma_e=1$) with different choices of $g, V, \Delta, J_c, J_v$. (c) OPCP amplitude versus fluence in the overdamped case with or without considering the depopulation time of the electrons, with the same model parameters as in (a).}
\end{figure*}

There is uncertainty in the determination of the electron decay rate ${\gamma}_e$. Recent theories have demonstrated that the electronic system can oscillate in an amplitude (Higgs) and phase (Goldstone) mode around the transient free energy minimum \cite{EcksteinPRL2017}, but there is no experimental evidence of such modes so far. A large $\gamma_e$ describes the overdamped case where the electronic subsystem adiabatically evolves into the transient free energy minimum, while a small $\gamma_e$ captures the underdamped case where $\Phi$ explores a larger region of the Mexican-hat potential via rapid oscillations of the Higgs and the Goldstone modes upon light excitation [Fig. S1(h)]. We demonstrate that the nature of the electronic decay, whether overdamped or underdamped, does not change our main finding, i.e. the reversal of the EI order and the concurrent anomalous phonon amplitude dependence. We simulated the aforementioned two cases using either ${\gamma}_e=1$, which typically characterizes an overdamped scenario with no oscillation [Fig. S1(a)-(d)], or ${\gamma}_e=0.1$, which describes the underdamped case [Fig. S1(3)-(h)]. With ${\gamma}_e=0.1$, the rapid Higgs/Goldstone oscillation clearly damps out in 0.25 ps. Despite the distinction in $\gamma_e$, the dynamics of $\Phi$ and $X$ are comparable qualitatively at times longer than 0.25 ps after the light excitation. The clear reversal when the pump fluence is higher than the critical fluence is realized independent of the value of ${\gamma}_e$. Further, alteration of the electron dephasing rate has a minimal affect on the reversal critical fluence [Fig. S2(a)]. Since an investigation of the behavior of the Higgs/Goldstone mode is beyond the scope of this work, we only simulate with the overdamped case hereafter.

\section{III. Difference between the calculated and experimental critical fluence}

In Section I, we defined $f{\triangleq} {\alpha}F\exp{(-\frac{4\ln{2}t^2}{\sigma^2})}$. To get the value of the simulated critical fluence $F_c$ in real units, we explicitly write out the scaling factor $\alpha=\frac{8Dq^2d^2(1-R)}{{\hbar}^2{\omega_{ph}}^2c\epsilon_0\sigma}$, where $D$ is the parameter defined in Section I, $q$ is electron charge, $d$ is the lattice constant, $R$ is the reflectance, $\hbar$ is the reduced Planck's constant, $\omega_{ph}$ is the pump light frequency, $c$ is the speed of light, $\epsilon_0$ is the vacuum permittivity, $\sigma$ is the pulse duration. Our model predicts that $F_c$ should be smaller when the pump polarization is parallel to the chain direction because $d$ is smaller and $R$ is larger in this geometry. This is consistent with the data reported in Ref. \cite{KaiserSciAdv2018}.  

We obtained a simulated $F_c\sim\:$6 mJ/cm$^2$ under the same reported experimental conditions \cite{KaiserSciAdv2018}, which needs to be scaled down by 16 times to match the real experimental $F_c$. There are several possible factors that give rise to this discrepancy. First, our model is an elementary model with a simplified two-band structure. There are also uncertainties in the determination of the parameters $J_c$, $J_v$, $\Delta$, $g$, and $V$. It is especially hard to determine $g$ and $V$ from experiment because they are not directly reflected in band structure measurements. Even though the chosen parameter values in Ref.\cite{EcksteinPRL2017,OhtaPRB2018} are claimed to reproduce the experimental data well in Ref. \cite{MonneyPRL2017}, we find the band dispersions displayed in several other ARPES papers differ subtly, giving rise to  an uncertainty in the  determination of the parameters \cite{ShinNCOMM2018,ZWTPRL2020,OkazakiARXIV2020,KingARXIV2019}. In Fig. S2(b) we show several OPCP amplitude versus pumping fluence curves obtained using different sets of parameter choices. It is clear that $F_c$ changes with the value of the microscopic parameters, but the qualitative trend stays the same. 

Second, $F_c$ is temperature-dependent. Increasing temperature makes the bandwidth and bandgap smaller and the Mexican-hat minimum shallower, both of which promote the order-parameter switch, thereby decreasing $F_c$. Our simulation corresponds to $T$ = 0 K case, and therefore gives an upper bound of $F_c$. As shown in Ref. \cite{KaiserSciAdv2018}, a 100 K increase in the temperature can make $F_c$ several times smaller. As such, it is possible that $F_c$ at finite temperature is much smaller than 0 K. To give an accurate estimate, a temperature dependent model needs to be considered. 

Third, transient heating effects are not accounted for in our model. For pump photon energies above the insulating gap, which is the case in our experiment, there will be finite absorption leading to transient heating of the electronic subsystem. This will contribute towards quenching the electronic potential in a similar manner to coherent electric field driving. Therefore neglecting transient heating effects will cause $F_c$ to be over-estimated. On the other hand, this approximation may be more realistic for the case of sub-gap pumping where absorption is suppressed.   

Taken altogether, these three factors lead us to conclude that our microscopic model and the corresponding simulation qualitatively reproduce the experimental observation. We also emphasize that regardless of the choice of parameters, the anomalous behavior of an OPCP is well reproduced in all simulations, demonstrating our conclusions are robust against the specific value of the parameters.

\section{IV. Ultrafast heating and subsequent cooling}
We note that ultrafast heating and subsequent cooling in both the electronic and structural channels upon pumping are not considered in our coherent-drive microscopic model. We demonstrate here that ignoring the heating and the subsequent cooling does not change the major conclusions. 

We first consider the electronic channel. Upon pumping, the electrons are excited into the conduction bands and the effective electronic temperature increases dramatically. The ultrafast heating of the electrons is accompanied by the transient restoration of higher symmetry, i.e. Mexican hat becomes parabolic, producing a qualitatively similar effect to that imparted by our coherent driving model. In both cases, one can deterministically engineer the final state through pumping. In our model we introduce light perturbation via a coherent Peierls phase, implying an immediate relaxation to lower symmetry as soon as the pulse excitation is over. Although this theoretical treatment is extensively utilized, in reality hot electrons will thermalize with the lattice through EPC with a characteristic depopulation time of 1 ps in Ta\textsubscript{2}NiSe\textsubscript{5} as measured by time-resolved optical and electron spectroscopy \cite{MonneyPRB2018, KaiserSciAdv2018, MonneyPRL2017, ShinNCOMM2018, ZWTPRL2020, OkazakiARXIV2020}. This implies the higher symmetry exists beyond the time duration of the pulse. To take this depopulation time into account, we simulate the pulse as a 1 ps exponential decay convolved with a Gaussian. The dynamics of the order parameters are very similar to the dynamics in the coherent quench case, except for the fact that the order parameters reach their stable states after longer time (1-2 ps) due to thermalization. We show the OPCP fluence dependence with thermalization in Fig. S2(c). Compared with the coherent quench case, the switch to the counterpart state occurs at a slightly lower fluence but the trend is the same. Therefore, we conclude here that the heating and cooling of the electrons do not influence our major conclusion that the switch is achievable by increasing fluence and observable via the OPCP fluence dependence. 

To estimate the lattice heating, we use the formula ${\Delta}T=\frac{(1-R)F}{C{\rho}{\delta}}$ to calculate the lattice effective temperature increase, where $R$ is the reflectance, $\rho$ is the density, $C$ is the specific heat capacity, and $\delta$ is the optical penetration depth for the pump photon energy (1 eV) \cite{BorisPRB2017,TakagiNCOMM2017}. Our experiments were performed at 80 K with a pump fluence of 0.5 mJ/cm\textsuperscript{2}. Using these values, we obtain a temperature increase of 15 K. Therefore, the lattice temperature is far below $T_c$ after pumping and the lattice temperature induced change of band structure is negligible. Thus, the lattice heating can also be ignored.

\section{V. Non-monotonic behavior of the OPCP amplitude versus fluence above the critical fluence}
As depicted in Fig. S2, the OPCP amplitude as a function of pump fluence shows a non-monotonic behavior when $F>F_c$. The amplitude of the ``oscillation" and the fluence where they emerge are dependent on the specific values of the microscopic parameters. This behavior arises from the strong feedback between the electronic and structural order parameter dynamics immediately after excitation. Because $\Phi$ always responds more rapidly than $X$, the subtle mismatch of the time when the two order parameters cross zero will influence the phonon amplitude. 

We take the case with overdamped dynamics as an example. In such a case, $\Phi$ relaxes into the potential minimum instantaneously, while $X$ takes a longer time to settle into the minimum depending on the phonon damping rate. When the pump fluence just surpasses $F_c$, once the pump excitation is over and $X$ starts to approach zero from the negative side, Re($\Phi$) does not cross zero but exhibits a partial regression back to the initial state [Fig. S1(a) and (c)]. This incomplete return to the initial state in turn tilts the phonon potential in the opposite direction as $X$ is evolving, thus exerting resistance to the phonon and decreasing the phonon amplitude. However, as the pump fluence further increases, $X$ crosses zero more quickly and the aforementioned temporal mismatch between $X$ and $\Phi$ crossing zero will be smaller, leading to a slight increase of the phonon amplitude. This explains the non-monotonic behavior that occurs just above $F_c$ in Fig S2(a). After careful examination of the dynamics of both channels at each pump fluence, we find that this temporal mismatch occurs twice as the pump fluence is increased, yielding the two ``dips" above $F_c$ in the overdamped cases [Fig. S2(a) and (b)]. The ``dips" finally disappear after the pump fluence is high enough so that $\Phi$ directly crosses zero without returning partially back to the initial state. Thereafter $X$ and $\Phi$ will not experience any mismatch and the phonon amplitude shows a smooth monotonic dependence on fluence. 

The non-monotonic behavior is stronger for the underdamped case because $\Phi$ undergoes Higgs/Goldstone oscillations [Fig. S2(a)]. The feedback between the two channels thus also lasts longer, creating more complicated dynamics. In addition to the mismatch between the time when $X$ and $\Phi$ cross zero as discussed in the overdamped case, the order parameters can now also oscillate back-and-forth between the minima on either side of zero [Fig. S1(e)-(h)]. This makes the final state more sensitive to pump fluence compared to the overdamped case. In other words, the order parameter is more susceptible to reversal upon small changes in pump fluence, leading to sharper modulations of the phonon amplitude. 

We also note that in the overdamped case where we account for electronic heating and cooling, increasing pump fluence will induce a greater number of reversals back to the initial state within one phonon period (0.5 ps), and the oscillations in Fig. S2(c) actually stem from these back-and-forth reversals. In contrast, the reversal only occurs once in the overdamped coherent pumping case. This discrepancy reveals that the long-time dynamics of the coupled system is strongly influenced by the dynamics immediately after the excitation. While these differences lead to quantitatively different long time behaviors, they do not alter the main conclusion of this work, i.e. the observation of the first reversal and the non-monotonic behavior of the OPCP amplitude versus fluence above $F_c$.

\section{VI. ISRS/DECP phonon simulation}

As mentioned in the main text, conventional Raman active phonon modes are launched through the impulsive stimulated Raman scattering (ISRS) or displacive excitation of coherent phonons (DECP) mechanisms. Several references have discussed and summarized the disparities between and the unification of these two mechanisms \cite{DresselhausPRB1992, StantonPRL1994, MerlinPRB2002, MisochkoPLA2011}. We simulated conventional ISRS/DECP phonons using the simplified formula from \cite{MisochkoPRB2016}:
\begin{equation}
{{\partial}_t}^2{X}=-{{\omega}_{ph}}^2X-2{\gamma}_{ph}{\partial}_tX+F(t)
\end{equation}
where in the displacive case, $F(t)=D\theta(t)$ and in the impulsive case $F(t)=F\delta(t)$. Convolving with the Gaussian-envelope pulse we have $F(t)=D\frac{1+\erf{(\sqrt{\frac{4\ln({2})}{\sigma^2}}t})}{2}$ for the DECP case and $F(t)=F\exp(-\frac{4\ln(2)t^2}{\sigma^2})$ for the ISRS case, where $D$ and $F$ are normalized fluences. The dynamics for both cases are generally the same except for a $\pi/2$ phase shift. As shown in Fig. S2(a), the amplitude of the ISRS/DECP-launched phonon has a linear fluence dependence.

\section{VII. Experiment and fitting details}

Single crystals of Ta\textsubscript{2}NiSe\textsubscript{5} were grown by chemical vapor transport reaction. First, a powder of Ta\textsubscript{2}NiSe\textsubscript{5} was synthesized by the solid-state reaction from a stoichiometric mixture of its elements. They were sealed in an evacuated quartz tube and heated at 900 \textcelsius{} for 5 hours. Next, the powder (around 2 g) and chunks of iodine (50 mg) were loaded in a sealed quartz tube, which was put in a two-zone furnace. The temperature for the growth sides were kept at 875 \textcelsius{} and 800 \textcelsius{} respectively for one month. The samples were cleaved along the (010) direction immediately before the experiment to obtain a fresh and smooth surface.

\begin{figure}[t]
\includegraphics[width=3.375in]{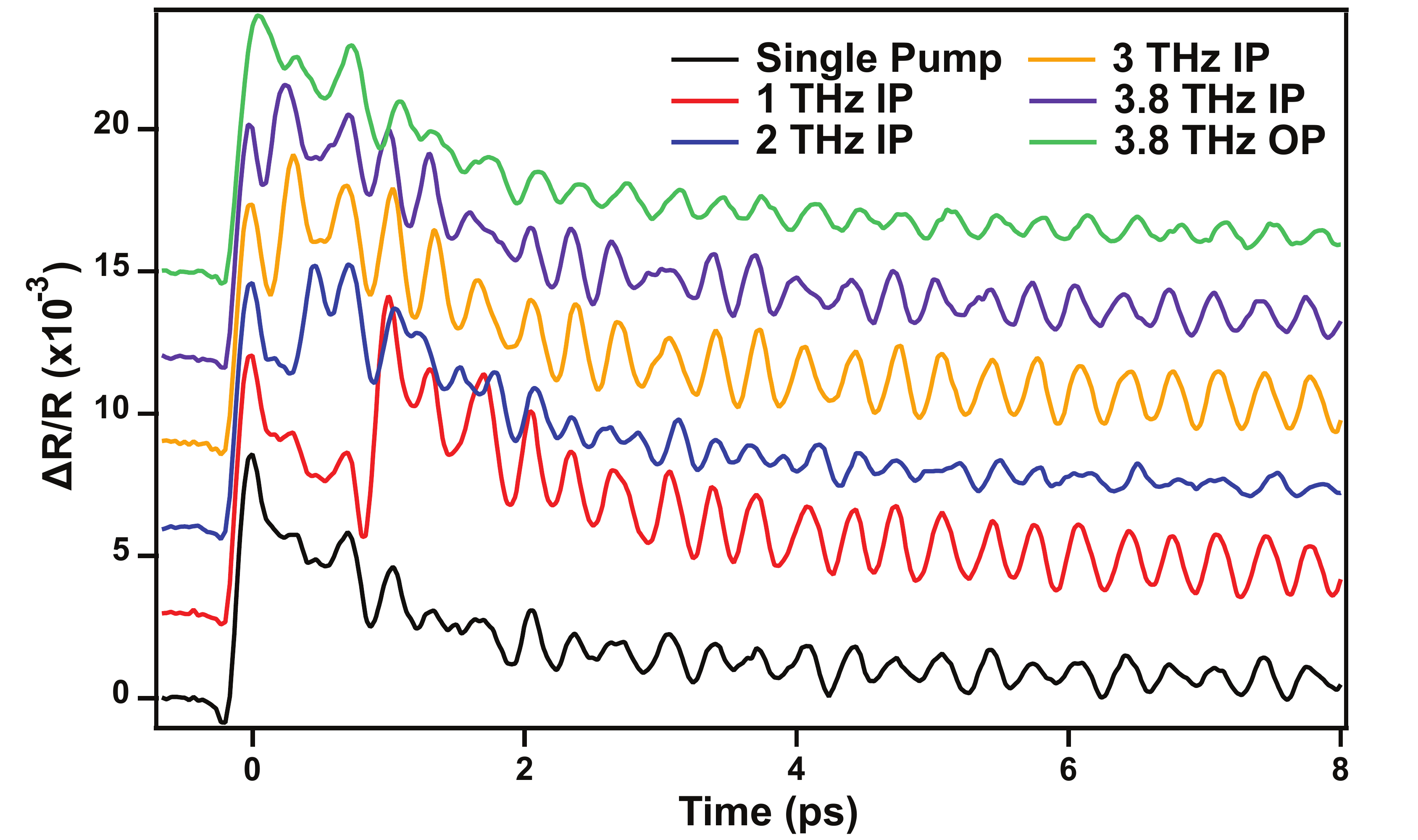}
\label{FigS3}
\caption{Transient reflectivity in response to both a single pump (black) and two pumps (other colors). In the double-pump data, the delays between the two pumps are set to be either in-phase (IP) or out-of-phase (OP) with a certain phonon, as indicated in the legend. Each curve is vertically offset for clarity.}
\end{figure}

For double-pump experiments, the sample temperature was fixed at 80 K. In the experimental setup, a Ti:sapphire amplified laser operating at 1 kHz produces 800 nm pulses with 40 fs time duration. A small portion of the power is used as the probe. The remainder seeds an optical parametric amplifier (OPA) and generates near infrared light tuned to 1200 nm with a duration of 80 fs, which is used for the the pump pulse(s). The fluence of each pump was set to $\sim$0.5 mJ/cm\textsuperscript{2}, which is around $F_c$ at 80 K, considering the temperature dependence of $F_c$ \cite{KaiserSciAdv2018}. Our simulation further substantiates that the fluence is slightly below $F_c$ as shown later. The polarizations of both pulses were set to be perpendicular to the (100) axis of the sample, since pumping with a parallel polarization was reported to generate phonons less efficiently \cite{KaiserSciAdv2018}.

\begin{figure}[t]
\includegraphics[width=3.375in]{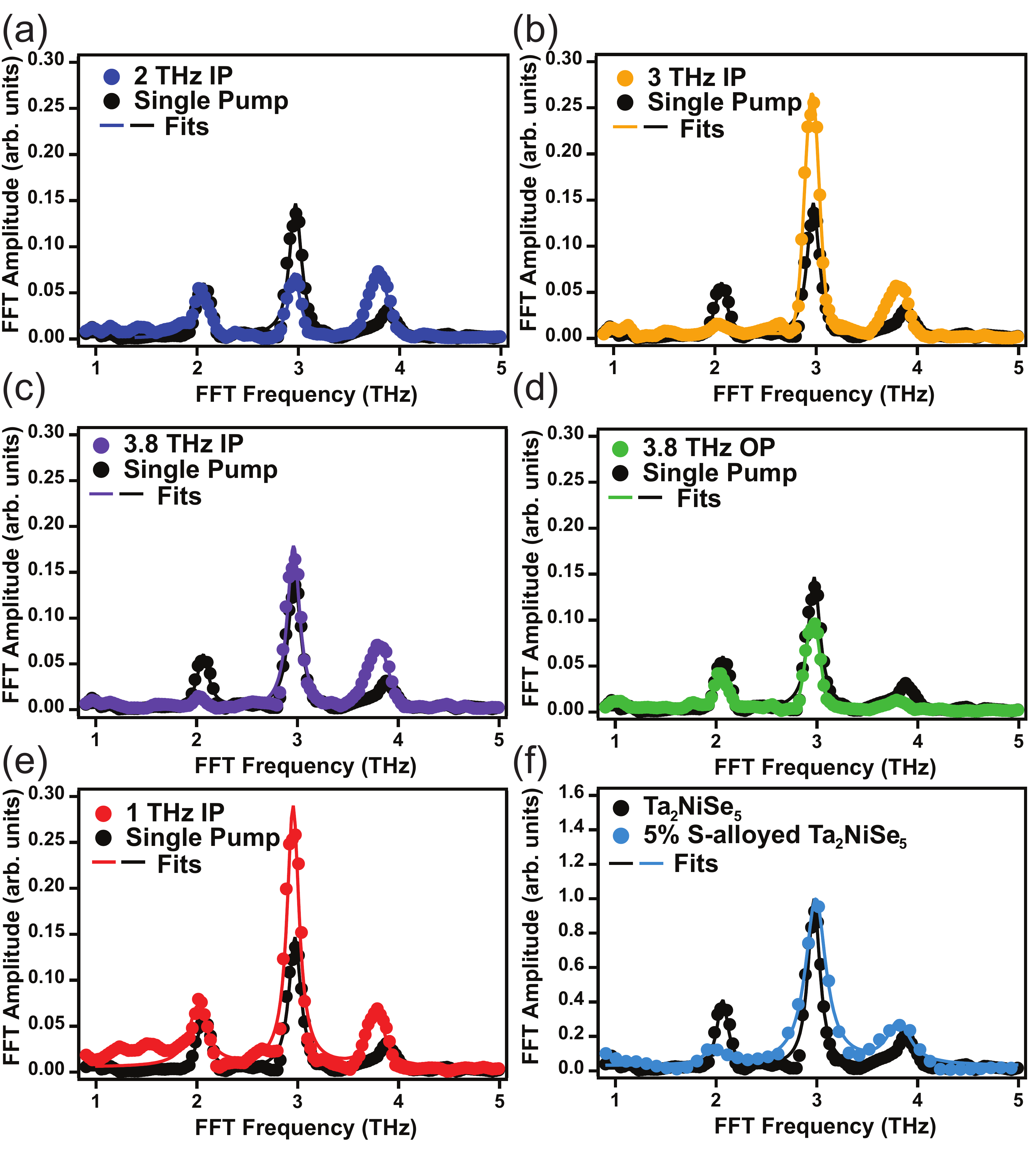}
\label{FigS4}
\caption{FFTs of experimental double- and single-pump transient reflectivity data taken on pristine and 5\% S-alloyed Ta\textsubscript{2}NiSe\textsubscript{5}. All FFTs are taken in the time interval between 0 ps and 8 ps. (a - e) FFT spectra of the traces shown in Fig. S3. The black dots correspond to the spectrum of the single pump excitation, shown in each panel as a reference, while the dots of other colors indicate the spectra of the double-pump data. The solid lines correspond to the Lorentzian fits. (f) Single-pump FFT spectrum for a 5\% S-alloyed Ta\textsubscript{2}NiSe\textsubscript{5} (cyan) with the data on the pristine sample as a reference (black).}
\end{figure}

Transient differential reflectivity curves with double- and single-pump are shown in Fig. S3. All curves exhibit a clear beat pattern. The quick rise upon the arrival of the pump and the ensuing exponential decay ($\sim$ 1 ps) characterize photocarrier generation and recombination respectively, in agreement with previous results \cite{KaiserSciAdv2018,MonneyPRB2018}. The beat pattern indicates the coexistence of multiple coherent phonons. Three phonons centered at around 2 THz, 3 THz and 3.8 THz are identified after taking the fast Fourier transforms (FFT) of the transient reflectivity data, as depicted in Fig. S4. The reported 1 and 4 THz phonons are missing \cite{OkazakiARXIV2020,ZWTPRL2020,MonneyPRB2018}.

To test whether we are able to amplify or suppress the different observed phonons, we excite the sample with two pump pulses with the time delay between them tuned to be either in-phase (IP) and out-of-phase (OP) with each phonon respectively. As such, there are six configurations in total. Note that when the time delay is IP with the 2 THz phonon, it is nearly OP with the 3 THz phonon. Similarly, when the time delay is IP with the 3.8 THz phonon, it is nearly OP with the 2 THz phonon. Therefore, two configurations are redundant and we executed the double-pump experiment using 4 different fixed delays between the two pumps. The corresponding double-pump FFT spectra are displayed in Fig. S4, together with the single pump FFT spectrum. In Fig. S4(b), it is demonstrated that when the time delay between the two pump pulses is resonant with the 3 THz phonon, the 3 THz phonon amplitude is amplified by almost two times. Simultaneously, the 3.8 THz phonon is slightly enhanced because this time delay is also partially IP with its period, while the 2 THz phonon is suppressed due to the nearly OP time delay. Similar analysis can be used to interpret the 3.8 THz IP and OP pumping cases in Fig. S4(c) and S4(d). However, an anomalous behavior is observed in the 2 THz IP pumping configuration as shown in Fig. S4(a). Although the 3 THz phonon is suppressed due to the OP time delay, and the 3.8 THz phonon is amplified due to nearly IP pumping, the 2 THz phonon is not enhanced.

\begin{figure}[t]
\includegraphics[width=3.375in]{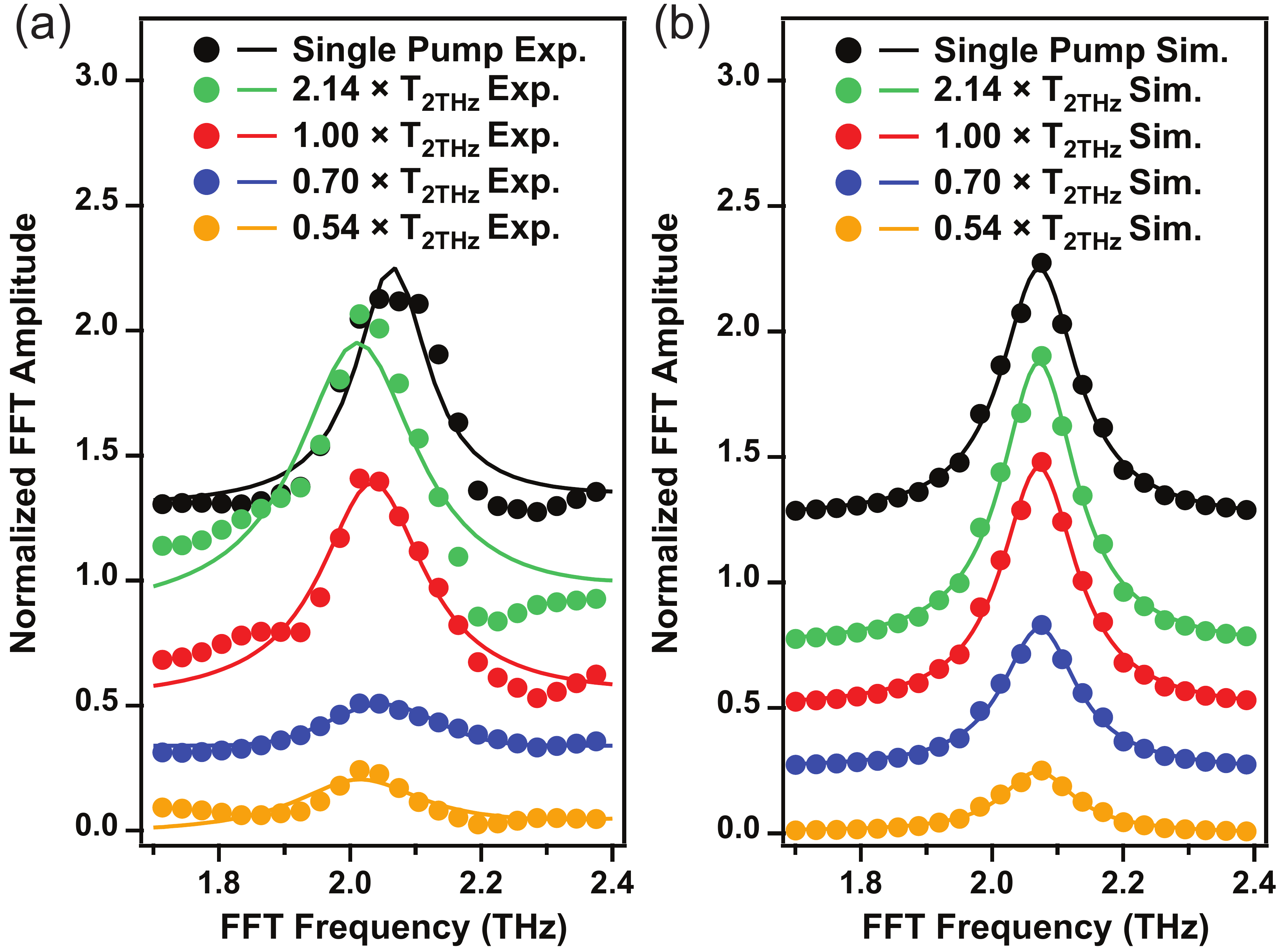}
\label{FigS5}
\caption{FFT spectra of the 2 THz phonon obtained from (a) experiment and (b) simulation of the microscopic model at time delays equal to 2.14$\:\times\:T_{2THz}$ (green, corresponds to the 1 THz IP pumping), 1$\:\times\:T_{2THz}$ (red, corresponds to the 2 THz IP pumping), 0.7$\:\times\:T_{2THz}$ (blue, corresponds to the 3 THz IP pumping), and 0.54$\:\times\:T_{2THz}$ (yellow, corresponds to the 3.8 THz IP pumping). The single pump results are displayed as a reference. Each curve is vertically offset for clarity and normalized by the peak value of the single pump FFT.}
\end{figure}

The reported 1 THz OPCP \cite{KaiserSciAdv2018} is missing in our single-pump spectrum. Fig. S4(e) displays the FFT spectrum of the 1 THz IP pumping. 1 THz IP pumping should be nearly IP with all the phonons, but only the 3 THz and 3.8 THz phonon are amplified. There is still no 1 THz phonon after pumping resonantly with it, and the enhancement of the 2 THz phonon is negligible, echoing the 2 THz IP double-pump results. Similar results were reproduced at different spots on two samples. This further demonstrates the 2 THz phonon is an OPCP unlike the 3 THz and 3.8 THz phonons.

We also conducted coherent phonon spectroscopy measurement on 5\% S-doped Ta\textsubscript{2}NiSe\textsubscript{5}. The FFT spectrum shows a much weaker 2 THz phonon than the undoped case, but similarly intense 3 THz and 3.8 THz phonons, as shown in Fig. S4(f). This observation is further evidence that the 2 THz phonon is coupled to the EI order while the 3 and 3.8 THz phonons are not, since S-doped Ta\textsubscript{2}NiSe\textsubscript{5} exhibits weaker EI order with a lower $T_c$ \cite{TakagiNCOMM2017}.

In principle we can fit the time traces with damped oscillations superposed atop an exponentially decaying background:
\begin{equation}\label{Eqn1}
\frac{{\Delta}R}{R}=A\exp{(-\frac{t}{\tau_0})}+C+\sum_{i}B_i\exp{(-\frac{t}{\tau_{ph,i}})}\cos{(\nu_it+\phi_i)}
\end{equation}
where $A$ denotes the electronic background amplitude due to photocarrier generation with a decay time $\tau_0$, and $C$ characterizes the long heat escape time. $B_i$, $\tau_{ph,i}$, $\nu_i$, $\phi_i$ are the amplitude, lifetime, frequency and phase of the $i$th phonon respectively. Here $i$ runs from 1 to 3 corresponding to 2, 3 and 3.8 THz phonons. 

Equivalently, we can also fit the peaks in the FFT spectrum. A damped oscillation in the time domain transforms into a Lorentzian in the frequency domain. Thus, we can fit the FFT data with three Lorentzians with the same definition of the corresponding parameters as defined above:
\begin{equation}\label{Eqn2}
\sum_{i}\frac{B_i}{(\nu-\nu_i)^2+(1/2\tau_{ph,i})^2}
\end{equation}

\begin{figure}[t]
\includegraphics[width=3.375in]{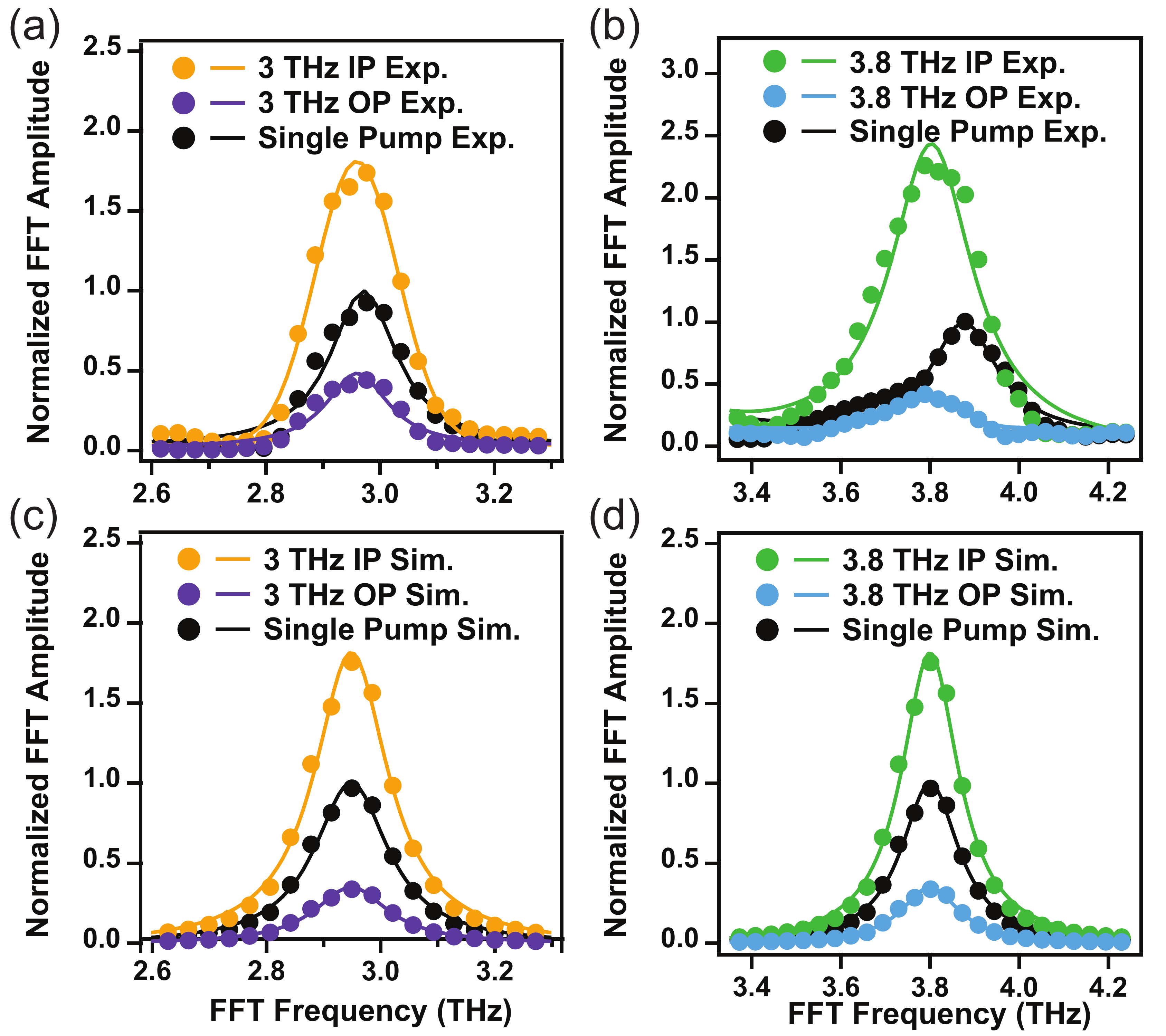}
\label{FigS6}
\caption{FFT spectra of the (a) 3 THz and (b) 3.8 THz phonons obtained from experimental data with IP and OP pumping. FFT spectra of the (c) 3 THz and (d) 3.8 THz phonons obtained from simulation results with IP and OP pumping. The single pump result is displayed in each panel (black) as a reference. Each curve is normalized by the peak value of the single pump FFT.}
\end{figure}

\section{VIII. Detailed comparison of double-pump experiment and simulation}

Here we present a detailed comparison between experiment and simulation of the complete pump time-delay dependence of the 2 THz phonon in the double-pump scheme. We zoom in on the 2 THz phonon FFT peak at the aforementioned five different time delays as shown in Fig. S5(a). Note that the time delays are expressed as multiples of the 2 THz phonon period ($T_{2THz}$). We simulated the time evolution of $X$ upon two-pulse excitation using the same time delays as in our double-pump experiment and obtained a FFT spectrum for each delay. The simulation results using a pump fluence $F=0.96F_c$ per pulse reproduce the experimental data well [Fig. S5(b)]. Simulation with pump fluence $F>F_c$ fails to reproduce the experimental data even qualitatively.

As a comparison, the 3 THz and 3.8 THz phonons exhibit enhancement with IP pumping and suppression with OP pumping, resembling the ISRS simulation very well [Fig. S6] and thus demonstrating their uncoupled nature. Also note that the measured frequency of all three phonons after double-pumping redshifts compared with the single-pump case due to higher net pumping fluences [Fig. S5 and Fig. S6]. This softening may come from carrier-excitation-induced lattice softening and phonon anharmonicity \cite{KurzPRL1995,FahyPRL1999,HasePRL2002}, which are ignored in our microscopic model and unrelated to the main results.